\documentclass[twocolumn,secnumarabic,amssymb, nobibnotes, aps, prd]{revtex4-1}

\newcommand{\env}[1]{\texttt{#1}}
\usepackage{graphicx,url,epstopdf,color}

 \definecolor{Black}{named}{Black}
 \definecolor{Blue}{named}{Blue}
 \definecolor{Red}{named}{Red}

\def\la{\mathrel{\mathpalette\fun <}}
\def\ga{\mathrel{\mathpalette\fun >}}
\def\fun#1#2{\lower3.6pt\vbox{\baselineskip0pt\lineskip.9pt
  \ialign{$\mathsurround=0pt#1\hfil##\hfil$\crcr#2\crcr\sim\crcr}}}

\begin{document}

\title{Method to Look for Imprints of Ultrahigh Energy Nuclei Sources}

\author{G.~Giacinti}
\email[]{giacinti@apc.univ-paris7.fr}
\affiliation{AstroParticle and Cosmology (APC, Paris), 10, rue Alice Domon et L\'eonie Duquet, 75205 Paris Cedex 13, France}

\author{D.~V.~Semikoz}
\email[]{dmitri.semikoz@apc.univ-paris7.fr}
\affiliation{AstroParticle and Cosmology (APC, Paris), 10, rue Alice Domon et L\'eonie Duquet, 75205 Paris Cedex 13, France}
\affiliation{Institute for Nuclear Research of the Russian Academy of Sciences, 60th October Anniversary prospect 7a, Moscow 117312, Russia}

\begin{abstract}
We propose a new method to search for heavy nuclei sources, on top of background, in the Ultra-High Energy Cosmic Ray data. We apply this method to the 69 events recently published by the Pierre Auger Collaboration~\cite{AugerPaper} and find a tail of events for which it reconstructs the source at a few degrees from the Virgo galaxy cluster. The reconstructed source is located at $\simeq8.5^{\circ}$ from M87. The probability to have such a cluster of events in some random background and reconstruct the source position in any direction of the sky is about $7\times10^{-3}$. The probability to reconstruct the source at less than 10$^{\circ}$ from M87 in a data set already containing such a cluster of events is about $4\times10^{-3}$. This may be a hint at the Virgo cluster as a bright ultra-high energy nuclei source. We investigate the ability of current and future experiments to validate or rule out this possibility, and discuss several alternative solutions which could explain the existing anisotropy in the Auger data.
\end{abstract}


\maketitle


\vspace{3pc}


\section{Introduction}
\label{Introduction}

The HiRes and Auger data have proved the existence of a cutoff at the highest energies in the Ultra-High Energy Cosmic Ray (UHECR) spectrum~\cite{spectrum_HiRes,auger_spectrum2008}. This indicates that UHECR sources are astrophysical. Due to the GZK cutoff or photo-disintegration, they must be located in the local Universe (at distances $r \la 100$\,Mpc). These astrophysical sources, which are still unknown, should be located in the Large Scale Structure (LSS) of matter.

The Pierre Auger Collaboration recently reported a shift towards a heavier composition in the Ultra-High Energy Cosmic Ray spectrum at the highest energies, above a few times $10^{19}$\,eV~\cite{Collaboration:2010yv}. The analysis of the Yakutsk EAS Array muon data is also in agreement with this observation~\cite{Glushkov:2007gd}. However, these results are still controversial: the measurements of HiRes experiment~\cite{BelzICRC} and some preliminary studies of the Telescope Array~\cite{TA} are consistent with a proton composition.

For the moment, methods to search for the sources of UHECRs have been presented for proton or light nuclei primaries. See for example Refs.~\cite{Golup:2009cv,Giacinti:2009fy}.

In case UHECR are heavy nuclei, looking for their sources would be a harder task: for example, 60\,EeV iron nuclei behave as $\simeq2$\,EeV protons in the Galactic Magnetic Field (GMF), due to their similar rigidities $E/Z$. Refs.~\cite{Harari:1999it,Harari:2000az,Harari:2000he,Yoshiguchi:2004kd,Giacinti:2010dk} studied the propagation in the GMF of particles with such low rigidities, while Refs.~\cite{Vorobiov:2009km,Takami:2009qz} discussed the effect of varying the UHECR composition on the correlation of Auger events with active galactic nuclei.

The GMF displays both a large scale and a random small scale structure, which are respectively known as the regular and the turbulent components.

In Ref.~\cite{Stanev:1996qj}, T.~Stanev suggested one of the first models of the regular GMF, describing analytically with logarithmic spirals the field structure in the Galactic disk. Other models of the disk field were formulated by D.~Harari~et~al.~\cite{Harari:1999it}, and by P.~Tinyakov and I.~Tkachev~\cite{Tinyakov:2001ir}. Then, M.~Prouza and R.~Smida built a model which adds a halo contribution made of toroidal and poloidal fields~\cite{PS,Kachelriess:2005qm}. The implications on the GMF modeling of recent rotation measure maps were reported by J.~L.~Han~et~al.~\cite{Han:2006ci,Han:2009jg,Han:2009ts}. Other models have been suggested for the disk field, such as a toroidal field consisting of concentric rings by J.~P.~Vall\'ee~et~al.~\cite{Vallee} and another axisymmetric field by L.~Page~et~al.~\cite{Page:2006hz}. Some recent models for the disk field also display a spiral pattern based on the structure of the NE2001 thermal electron density model (J.~C.~Brown~et~al.~\cite{Brown:2007qv}) or on the spiral structure of the Milky Way (Y.~Y.~Jiang~et~al.~\cite{Jiang:2010yc}) deduced from HII regions and giant molecular clouds~\cite{Hou:2009gn}. X.~H.~Sun~et~al. proposed several GMF models and confronted them with the data~\cite{Sun:2007mx,Sun:2010sm}. However, currently no theoretical GMF model can reasonably well fit all experimental data, as Refs.~\cite{Jansson:2009ip,Waelkens:2008gp} show.

References~\cite{Harari:2002dy,Tinyakov:2004pw} discuss the turbulent component modeling and Ref.~\cite{Han:2004aa} investigates its spectrum. Its implications on the propagation of UHECRs are studied in Refs.~\cite{Harari:2002dy,Tinyakov:2004pw,TurbGMFiron}.

In many cases, one may not detect the signatures from Ultra-High Energy (UHE) nuclei sources without a more precise knowledge on the Galactic Magnetic Field than currently available~\cite{Giacinti:2010dk}. Astronomy with UHE heavy nuclei can look very different from astronomy with light primaries. In particular, multiple images of the same source can appear even at the highest energies. The images of nearby galaxy clusters in some recent GMF models have been shown in Ref.~\cite{Giacinti:2010dk}, illustrating the challenges of heavy nuclei astronomy.

However, the images of some UHE nuclei sources could be detectable in favorable cases, without requiring an improved knowledge of the GMF. We propose in this paper a method to look for heavy nuclei sources, in such situations. It is an extension to the case of heavy nuclei of the method we presented in Ref.~\cite{Giacinti:2009fy} for protons and light nuclei.

We apply the method to the list of 69 events with energies $E \geq 55$\,EeV recently published by the Pierre Auger Collaboration~\cite{AugerPaper}. We find in this data a cluster of events for which the reconstructed source lies near Virgo, which is in line with the supposition of Ref.~\cite{DSPaper}. We generate sets of 69 background events following the exposure and spectrum of Auger. The probability to have such a cluster of events and reconstruct the source position in any direction of the sky is about $7\times10^{-3}$. Assuming that the cluster already exists in the data due to another reason, we study how often the method would reconstruct the source at less than 10$^{\circ}$ from M87. This probability is about $4\times10^{-3}$. The combined probability of having such a cluster and reconstructing the source position at less than 10$^{\circ}$ from M87 is about $3\times10^{-5}$. Being the largest galaxy cluster in the local universe and hosting the powerful active galaxy M87, the Virgo cluster is, theoretically, a good candidate for being home of one or several source(s) of UHECR.

Nevertheless, both the ``limited'' amount of data and the poor knowledge of the GMF global geometry prevent us to conclude firmly whether the detected tail of events really comes from Virgo or not. We analyse the ability of current and next generation experiments to test this possibility. We also review alternative solutions which could explain the Auger data.

This paper is structured as follows: In Section~\ref{Method}, we present a method to search for detectable imprints of UHE nuclei sources. We apply it to the recently published Auger data, in Section~\ref{Data}. We find with this method a signal which may hint at the Virgo cluster as a bright UHE nuclei source. This result is tested with a ``blind-like'' analysis in Section~\ref{BlindAnalysis}, by analysing successively the lists of 27 and $69-27=42$ events recorded by Auger. In Section~\ref{Discussion}, we present a detailed and critical analysis of the possibility that Virgo may be a UHE nuclei source. We also discuss alternative reasons which could explain the anisotropy in the Auger data.


\section{Method to search for heavy nuclei sources}
\label{Method}

At the highest energies, for a source of UHE protons or light nuclei located far enough from the Galactic plane, the GMF approximately shifts its events in a sector-shaped region on the celestial sphere, on one side of the source~\cite{Giacinti:2009fy}. The vertex of the sector is, theoretically, located at the source position, and its opening angle depends on the ratio of deflections due to the turbulent and regular components of the GMF. In a first approximation, the angular distance between the source and an event of energy $E$ is proportional to $1/E$~\cite{Golup:2009cv}. The associated proportionality factor will be called the ``deflection power'' of the regular GMF, $\mathcal{D}$, in the following.

Contrary to the naive idea that sources of heavy nuclei would display the same features only enlarged by a factor $Z$, Refs.~\cite{Harari:1999it,Harari:2000he,Harari:2000az,Giacinti:2010dk} point out that their images often have more complicated shapes, even at energies $E\geq60$\,EeV. For instance, sources can have several images. They can appear above or below a given energy threshold, and merge into one single image when the energy is increased. They can also be strongly distorted on the celestial sphere and display an energy ordering far from the $1/E$ behavior expected close to the ballistic regime -see details in Ref.~\cite{Giacinti:2010dk}. Moreover, these images are very dependant on the considered model of GMF. Hence, for nuclei sources, a better knowledge than currently available of the GMF geometry would be needed in many cases, in order to find out an efficient and particular algorithm to detect their events and reconstruct their positions.

\begin{figure}
\begin{center}
\includegraphics[width=0.49\textwidth]{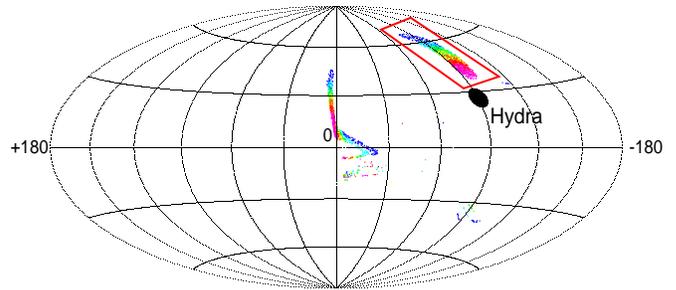}
\end{center}
\caption{Arrival directions of ultra-high energy iron nuclei emitted by the Hydra galaxy cluster, in Galactic coordinates, and deflected in the PS regular GMF model. Colors correspond to the energies of the emitted nuclei: dark blue stands for 60\,EeV nuclei, light blue: 70\,EeV, green: 80\,EeV, yellow: 90\,EeV, orange: 100\,EeV, red: 120\,EeV, and magenta: 140\,EeV. The Hydra cluster is represented by the black disk. The image which is approximately ``enlarged proton-like'' is surrounded by the red box.}
\label{ImagesGC}
\end{figure}

Meanwhile, one can still try to find simple proton-like orderings, with $\sim Z$ times larger angular scales. As shown below, this can indeed happen in some favorable specific cases. For some types of GMF models and for some positions on the sky, at least one image of an iron source can look like a more or less roughly enlarged proton-like image at high enough energies ($E \ga 50-60$\,EeV). We checked this point by computing the iron images of nearby galaxy clusters in the three recent models of the regular GMF which are considered in Ref.~\cite{Giacinti:2010dk}. One example of roughly ``enlarged proton-like'' image is the main image of Hydra cluster in the Prouza and Smida (PS) model~\cite{Kachelriess:2005qm,PS}, shown in Fig.~\ref{ImagesGC}. This image is surrounded by a red box. The hydra galaxy cluster is represented by the black disk, and colors correspond to the energies of the emitted nuclei: dark blue stands for 60\,EeV nuclei, light blue: 70\,EeV, green: 80\,EeV, yellow: 90\,EeV, orange: 100\,EeV, red: 120\,EeV, and magenta: 140\,EeV.

For this paper, we will focus on the favorable case of approximately ``enlarged proton-like'' images of sources. We leave the more frequent but more complicated cases for future works.

\begin{figure}
\centering
\includegraphics[width=0.49\textwidth]{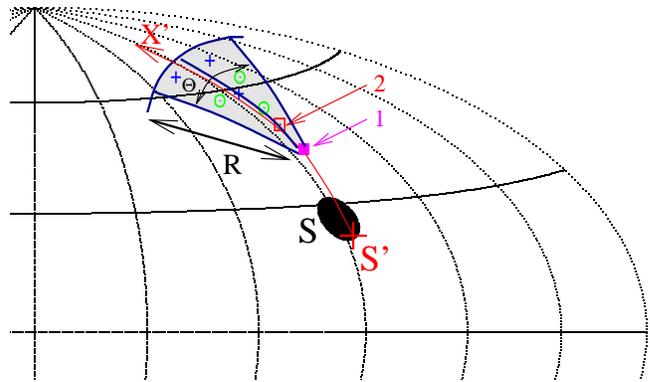}
\caption{Example of a UHE heavy nuclei source detection, in case of an ``enlarged proton-like'' image. The extended source S is shown by the black disk, and its cosmic rays are represented by decreasing energy order, by the magenta filled and red open boxes, green open circles and blue crosses. The ``sector'', with an opening angle $\Theta$ and extension $R$, is highlighted in grey -see text for details. ``1'' and ``2'' respectively point at the highest energy event and at the selected event in the list satisfying Eqn.~(\ref{shEe}). $X'$ axis is drawn in red. S' denotes the reconstructed source.}
\label{Fig_Method}
\end{figure}

The method we proposed to look for proton and light nuclei sources is presented in details in Section 3.1 of Ref.~\cite{Giacinti:2009fy}. In this work, we slightly modify this method in order to optimize it to the search for heavy nuclei sources. Since we want to scan over all the free parameters of the method, we try to have as few parameters as reasonably possible. There are four of them. A schematic image of this method is drawn in Fig.~\ref{Fig_Method}. The source S is represented by a black disk.

Our procedure starts selecting an event with energy $E_{1}\geq10^{20}$\,eV. We will call it the ``highest energy event''. It is denoted by ``1'' in Fig.~\ref{Fig_Method}.

Let us consider such an event. We do an assumption on the typical value of the local regular GMF deflection power, $\mathcal{D}$, and only consider the events which angular distance to the highest energy event, $d$, satisfies
\begin{equation}
d \leq R = \frac{\mathcal{D}}{55\mbox{EeV}} - \frac{\mathcal{D}}{E_{1}}\,.
\label{radius}
\end{equation}
The next step is to search for the events which energy $E_{2}$ and distance $d$ also fulfill the condition
\begin{equation}
d \leq \frac{\mathcal{D}}{E_{2}} - \frac{\mathcal{D}}{E_{1}}\,.
\label{shEe}
\end{equation}
The events satisfying this latter condition are tested one after another, by decreasing energy order, with the procedure described below.

Let us start with the event which has the highest energy among them. It is denoted by ``2'' in Fig.~\ref{Fig_Method}.

In the following, we focus on the events located in a given sector-shaped region of the sky. Its direction is defined by this event ``2''. This region is highlighted in grey in Fig.~\ref{Fig_Method}. We define it as an extension to spherical geometry, of a circular sector which vertex is located at the position of the highest energy event (``1'' in Fig.~\ref{Fig_Method}). More precisely, such region is the sub-region of a spherical lune with the highest energy event on one vertex. This sub-region contains the points of the spherical lune which angular distance to this vertex satisfy Eqn.~(\ref{radius}). We will refer to this region of the sky as the ``sector'' in this paper. Its opening angle is given by the second free parameter of the method, $\Theta$, and its extension by the angular distance $R$ -see Eqn.~(\ref{radius}). Its central axis, which divides the opening angle in two equal parts, is defined by the line containing both the events ``1'' and ``2''.

Let us define the correlation coefficient $Corr(X',1/E)$ for the events in the sector, where $E$ denotes their energies and $X'$ their angular distances to the vertex of the sector. $X'$ is represented by the red axis in Fig.~\ref{Fig_Method}. The two last free parameters of the method are the minimal number $\mathcal{N}$ of events in the sector and the minimal value $C_{\min}$ of the correlation coefficient: If there are more than $\mathcal{N}$ events in the sector, and that $Corr(X',1/E) \geq C_{\min}$, there is a detection. Otherwise, the second event in the ordered list of events satisfying Eqn.~(\ref{shEe}) is tested. The procedure continues until either there is one detection, or all the events in the list are tested. In the latter case, there is no detection.

In case of detection, we reconstruct the source position as depicted in Section 3.3 of Ref.~\cite{Giacinti:2009fy}. The source is reconstructed along the $X'$ axis. This axis is not the exact central line of the sector. It is the axis which contains both the center of mass of all cosmic rays in the sector and the vertex of the sector. The position of the reconstructed source is given by the fit of $1/E$ versus $X'$. It is represented by a thick red cross, S', in Fig.~\ref{Fig_Method}.

For heavy nuclei with energies $E \geq 55$\,EeV, deflections on the celestial sphere can easily reach several tens of degrees. Even in good cases, this often leads to strong deviations to the linear shape of images and to the $1/E$ ordering of events. That is why, one should not expect for heavy nuclei sources the same excellent precision on the reconstruction of the source position as for proton sources.

To summarize, the method used in this paper has 4 free parameters:
\begin{enumerate}
 \item The deflection power $\mathcal{D}$, and the opening angle $\Theta$. The best value for $\mathcal{D}$ is mostly related to the strength of the regular GMF. The best value for $\Theta$ mostly depends on the ratio of deflections in the turbulent and regular components of the GMF. While the values of these contributions are not precisely known due to the lack of knowledge on the GMF, their most probable ranges can be inferred from the literature.
 \item The minimum number of events in the sector, $\mathcal{N}$, and the minimum value of the correlation coefficient $Corr(1/E,X')$ for the events in the sector, $C_{\min}$. Below these values, the considered features are rejected by the method.
\end{enumerate}


\section{Application to the data of the Pierre Auger Observatory}
\label{Data}

\begin{figure}
\begin{center}
\includegraphics[width=0.49\textwidth]{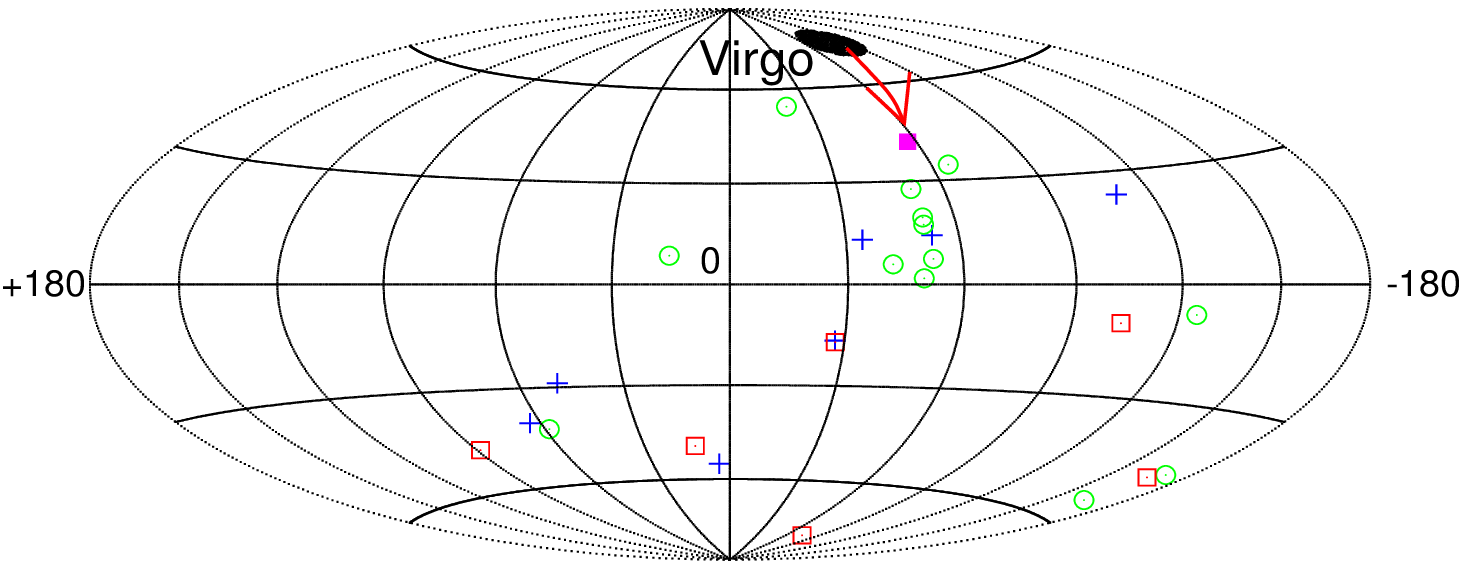}
\includegraphics[width=0.49\textwidth]{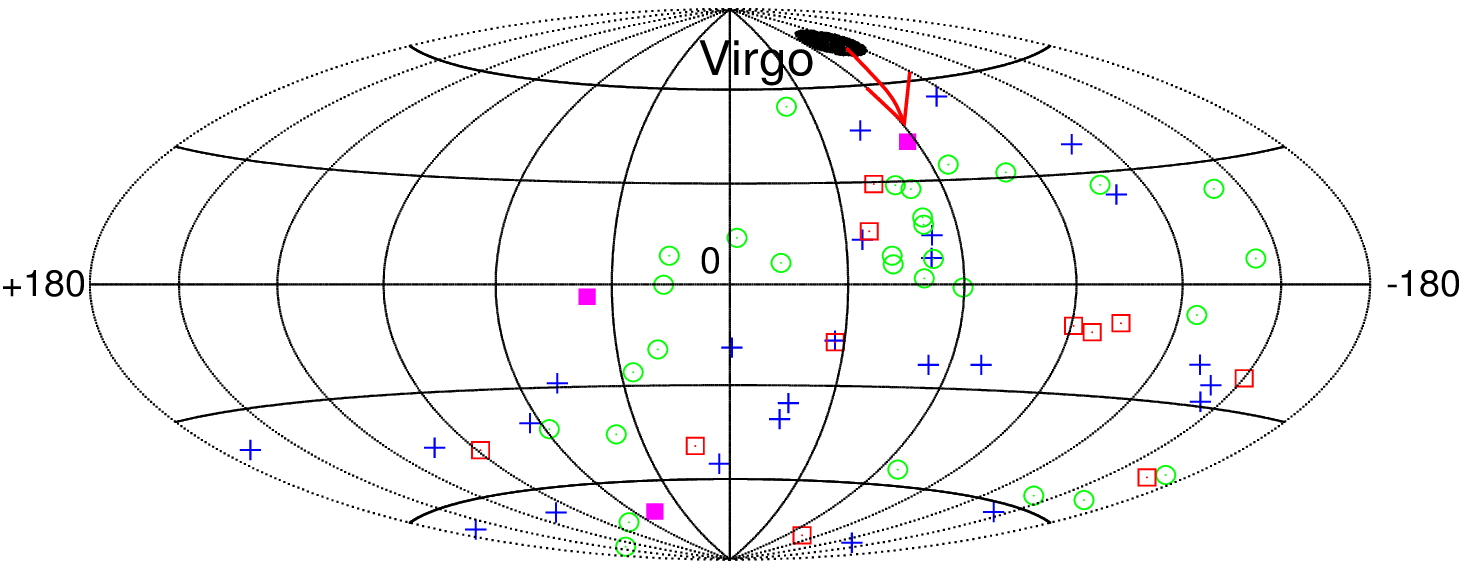}
\end{center}
\caption{Images, in Galactic coordinates, of Auger events with energies $E \geq 55$\,EeV, published in Ref.~\cite{AugerPaper}. \textbf{Upper panel:} First data set of 27 events~\cite{Cronin:2007zz,Abraham:2007si}, with the renormalized coordinates of Ref.~\cite{AugerPaper}. \textbf{Lower panel:} Full data set of 69 events. Black disk for the position on the sky of the Virgo cluster. Filled magenta boxes stand for events with energies $E \geq 10^{20}$\,eV, red open boxes for $10^{19.9}$\,eV\,$\leq E \leq 10^{20}$\,eV, green open circles for $10^{19.8}$\,eV\,$\leq E \leq 10^{19.9}$\,eV and blue crosses for $55$\,EeV\,$\leq E \leq 10^{19.8}$\,eV. The red arrows show the direction along which events emitted by Virgo would be shifted in the GMF if Cen A region events are nuclei from Virgo.}
\label{AugerEvents}
\end{figure}

The Pierre Auger Collaboration recently released in Ref.~\cite{AugerPaper} a list of 69 events with energies higher than 55\,EeV. They were recorded with a total integrated exposure of 20,370~km$^{2}\cdot$sr$\cdot$yr. In Fig.~\ref{AugerEvents} (upper panel), we plot in Galactic coordinates the positions of the first 27 events. They correspond to the first released data set of Refs.~\cite{Cronin:2007zz,Abraham:2007si}, renormalized as in~\cite{AugerPaper}. We plot in Fig.~\ref{AugerEvents} (lower panel), all the 69 events (current data set). Events with energies $E \geq 10^{20}$\,eV, $10^{19.9}$\,eV\,$\leq E \leq 10^{20}$\,eV, $10^{19.8}$\,eV\,$\leq E \leq 10^{19.9}$\,eV and $55$\,EeV\,$\leq E \leq 10^{19.8}$\,eV are respectively represented by filled magenta boxes, red open boxes, green open circles and blue crosses.

The most visible feature in the Auger data is an overdensity of events in the region $-60^{\circ}\la l \la-30^{\circ}$ and $0^{\circ}\la b \la30^{\circ}$. It was first discussed for the data set of 27 events in Ref.~\cite{Gorbunov:2008ef}. For the 69 events data set, it was studied in Ref.~\cite{AugerPaper}. In the following, we will call this part of the sky the ``Cen A region''. An important point was noted in Ref.~\cite{DSPaper}. It shows that if one excludes this overdensity, the rest of the sky could still be compatible with isotropy. The significance of the overdensity was computed both in Refs.~\cite{AugerPaper} and~\cite{DSPaper}. Ref.~\cite{AugerPaper} found 4\% with a Kolmogorov-Smirnov test, and Ref.~\cite{DSPaper} 3\% (resp. 2\%) with 3-point (resp. 4-point) autocorrelation functions.

We shall now analyse the set of 69 events with the method presented in the previous section. We scan the Auger data over all possible combinations of the 4 parameters of the model. We use discretized sets of values for each parameter.

We take $\mathcal{N} \in \left\lbrace 4,5,...,69\right\rbrace $. For the minimum value of the correlation coefficient, we take a set with steps equally spaced by 0.1: $C_{\min} \in \left\lbrace -1,-0.9,...,0.9\right\rbrace $. Existing measurements of the GMF allow to give typical estimates of $\simeq1-2.5^{\circ}$ deflections on the sky for $10^{20}$\,eV protons~\cite{Kachelriess:2005qm}. Then, assuming the heavy nuclei primaries to be iron ($Z=26$), we take for this analysis $\mathcal{D} \in \left\lbrace 26^{\circ}, 39^{\circ}, 52^{\circ}, 65^{\circ}\right\rbrace  \times 10^{20} \mbox{eV}$. According to the results of Ref.~\cite{Tinyakov:2004pw} on the relative contributions of the regular and turbulent components to the UHECR deflections, $\Theta \leq 80^{\circ}$ should be sufficient. Then, we take $\Theta \in \left\lbrace 10^{\circ},20^{\circ},...,80^{\circ}\right\rbrace $.

We confront below the data with $4.7\times10^{7}$ Monte Carlo simulations of random background. This background is made of exactly 69 events and follows the exposure of Auger experiment, \textit{i.e.} its local density statistically follows the exposure. Its energy spectrum follows the spectrum of the 69 Auger events, distributed in four logarithmically spaced bins: 55\,EeV-10$^{19.8}$\,eV, 10$^{19.8}$\,eV-10$^{19.9}$\,eV, 10$^{19.9}$\,eV-10$^{20}$\,eV and above 10$^{20}$\,eV. The exact distribution of energies within each of the first three bins does not significantly change the results below. The spectrum above $10^{20}$\,eV is poorly known, and Auger has only recorded 3 events at such energies. We take here for the events in the bin $E \geq 10^{20}$\,eV a $E^{-4.3}$ spectrum. This spectrum was proposed in Ref.~\cite{Abraham:2010mj} for events with $E \ga 10^{19.5}$\,eV. The value for the maximum observable energy, $E_{\max}$, is chosen according to results on propagation of nuclei. The figure 1 of Ref.~\cite{Allard:2005ha} shows that the iron nuclei propagation length rapidly falls below a few Mpc for energies above $3\times10^{20}$\,eV. Therefore, we take $E_{\max}=10^{20.5}\simeq3\times10^{20}$\,eV. We checked that taking lower values for $E_{\max}$ would only increase the significance of the signal detected below.

\begin{figure}
\begin{center}
\includegraphics[width=0.49\textwidth]{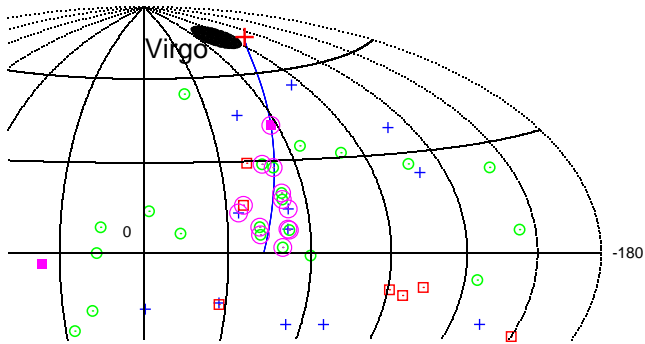}
\end{center}
\caption{Portion of the celestial sphere with the Auger data in Galactic coordinates, for the case of the best sector with $\Theta=40^{\circ}$ and $\mathcal{D}=39^{\circ} \times 10^{20}$\,eV. The 13 events contained in the sector are surrounded by magenta circles. The blue line represents the X' axis and the thick red cross, the position of the reconstructed source ($l \simeq -106^{\circ}$, $b \simeq 72.5^{\circ}$). The Virgo galaxy cluster is denoted by the black disk. Same color code for the Auger cosmic rays, as in Fig.~\ref{AugerEvents}.}
\label{Sector40degrees}
\end{figure}

For one of the three events with energies above 10$^{20}$\,eV, we find an interesting signal which is shown in Fig.~\ref{Sector40degrees}. The coordinates of this event are : $E=142$\,EeV, $l=-57.2^{\circ}$, $b=41.8^{\circ}$. It is located at $\simeq34^{\circ}$ from the center of Virgo. It plays the role of the ``highest energy event'' in the method. The best configuration is obtained for $\mathcal{N}=13$, $C_{\min}=0.6$, $\mathcal{D}=39^{\circ} \times 10^{20}$\,eV and $\Theta=40^{\circ}$. Among all possible configurations in the data, it is the one which has the lowest probability to be reproduced by the background. The value for $Corr(X',1/E)$ is $\simeq0.66$.

We computed the probability to obtain an at least as good feature in the background. Out of $4.7\times10^{7}$ generated sky maps of background, we found 3868 of such features for these 4 fixed parameters. This number has to be penalized over all possible values in the ranges of the 4 parameters of the method. As pointed out in Ref.~\cite{Finley:2003ur}, one cannot know \textit{a priori} the best values of the scan parameters. Therefore, one has to take into account any configuration in the background which probability is lower or equal to the probability of the specific feature detected in the data for $\mathcal{N}=13$, $C_{\min}=0.6$, $\mathcal{D}=39^{\circ} \times 10^{20}$\,eV and $\Theta=40^{\circ}$. In this study, for each value of ($\mathcal{D}, \Theta$) we scan all Monte Carlo realizations of the background over all values of ($\mathcal{N},C_{\min}$). We count all cases for which the probability to have a detection with a given value of ($\mathcal{N},C_{\min}$) is lower or equal to the probability of the best sector in the data, $3868/(4.7\times10^{7})$. After summing over all values of ($\mathcal{D}, \Theta$), the total number of such cases in the background is 311481. With our method, the probability of the feature is then: P$_{feature}\simeq6.6\times10^{-3}$.

For this configuration, the reconstructed source is located at $\simeq8.5^{\circ}$ from M87, at ($l \simeq -106^{\circ}$, $b \simeq 72.5^{\circ}$). It is near the boundaries of the Virgo cluster, which has an apparent radius of $\simeq5^{\circ}$ on the sky.

The position of the reconstructed source is drawn from the central value of the fit of $1/E$ versus $X'$. The uncertainties due to the fit are of the order of $\sim10^{\circ}$, because of the low energy ordering of events in the Cen A region. Moreover, even if M87 would be the only source in Virgo, the magnetic fields in the cluster can be sufficient to significantly deflect trajectories of UHE heavy nuclei inside and make shine the whole cluster as an extended source~\cite{virgoMF}.

The reconstructed position is then compatible with the Virgo cluster (or M87) being the source of (most of) the considered 13 events. In Fig.~\ref{Sector40degrees}, these 13 events are surrounded by magenta circles and the reconstructed source position is denoted by the thick red cross.

If we also add the constraint that the reconstructed source should be located at less than 10$^{\circ}$ from M87, the number of such cases in the background falls to 15 and 1214, respectively before and after the penalization. This corresponds to the following probability: P$_{feature,d(M87)<10^{\circ}}\simeq3\times10^{-5}$. The Virgo galaxy cluster was suggested to be a source of UHE nuclei by Ref.~\cite{DSPaper}. The events in the Cen A region and the 142\,EeV event were supposed to be its image, which is in line with the results shown in Fig.~\ref{Sector40degrees}.

However, one could argue that the overdensity can be explained by another reason than events emitted by the Virgo cluster. The Cen A region overdensity may for example be explained by a magnetic lensing effect, and the presence of the nearby event with $E \geq 10^{20}$\,eV may have triggered artificially the detection. The relevant value is then the probability to reconstruct the source at less than 10$^{\circ}$ from M87, in case one has already such a feature in the data (\textit{i.e.} at least 13 events, with a correlation coefficient $\geq0.6$). This is estimated by: P$_{d(M87)<10^{\circ}}=$\,P$_{feature,d(M87)<10^{\circ}}/$P$_{feature}\simeq4\times10^{-3}$. Thus, the probability that the source is reconstructed near Virgo due to a random fluctuation is \textbf{$\simeq$\,0.4\%}.

\begin{figure}
\begin{center}
\includegraphics[width=0.49\textwidth]{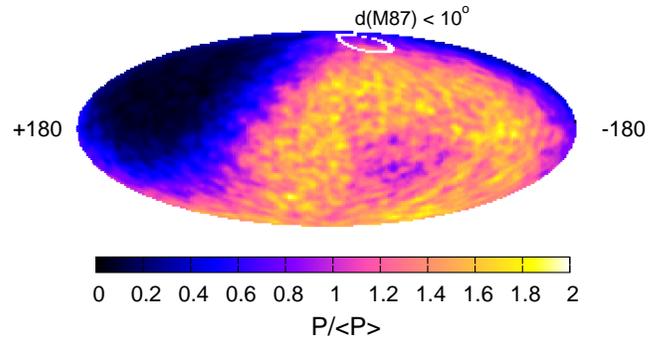}
\end{center}
\caption{Renormalized probability $P/<P>$ to reconstruct the source in any direction of the celestial sphere with random background following the Auger exposure and spectrum. $<~P~>$ represents the mean probability $P$ averaged over all bins of the sky map. The white circle indicates the $10^{\circ}$ region around M87.}
\label{AllRecSPositions}
\end{figure}

It is slightly lower than the value it would take in case of a random position of the reconstructed source. With the background events, the exposure of Auger favours the reconstruction of sources within its region of high exposures. In Fig.~\ref{AllRecSPositions} we plot the probability $P$ to reconstruct the source position in any direction of the sky for random background events, renormalized to $P/<P>$. $<P>$ is the mean probability to reconstruct the source in a given bin of the plot, averaged over all directions of the celestial sphere. All sectors with a probability of occurrence lower or equal to the probability of the best sector in the data have been considered. Therefore, the sum of all bins of Fig.~\ref{AllRecSPositions} adds up to $\sum P = P_{feature} \simeq 6.6 \times 10^{-3}$. The part of the sky where $P \ga (0.4-0.5) \times <P>$ globally corresponds to the directions in which Auger exposure is non-zero. The probability to reconstruct the source in regions which Auger is blind to is lower, though non-zero. The region of maximum probability, $P \simeq (1.6-1.9) \times <P>$, is a circular band within regions of high exposures. The white circle surrounds the part of the sky within $10^{\circ}$ from M87. In most of it, $P \sim (0.5-1) \times <P>$ (blue and purple colors). $P$ is slightly larger than $<P>$ in the smaller pink subregion at lower $b$ and larger $l$. On average, $P$ is slightly lower than $<P>$ in the $10^{\circ}$ radius region.

It may be noteworthy to point out the main two other possible sector angles $\Theta$ that one can consider when analysing the Auger data. The value of $\mathcal{D}=39^{\circ} \times 10^{20}$\,eV is left unchanged:

\begin{figure}
\begin{center}
\includegraphics[width=0.49\textwidth]{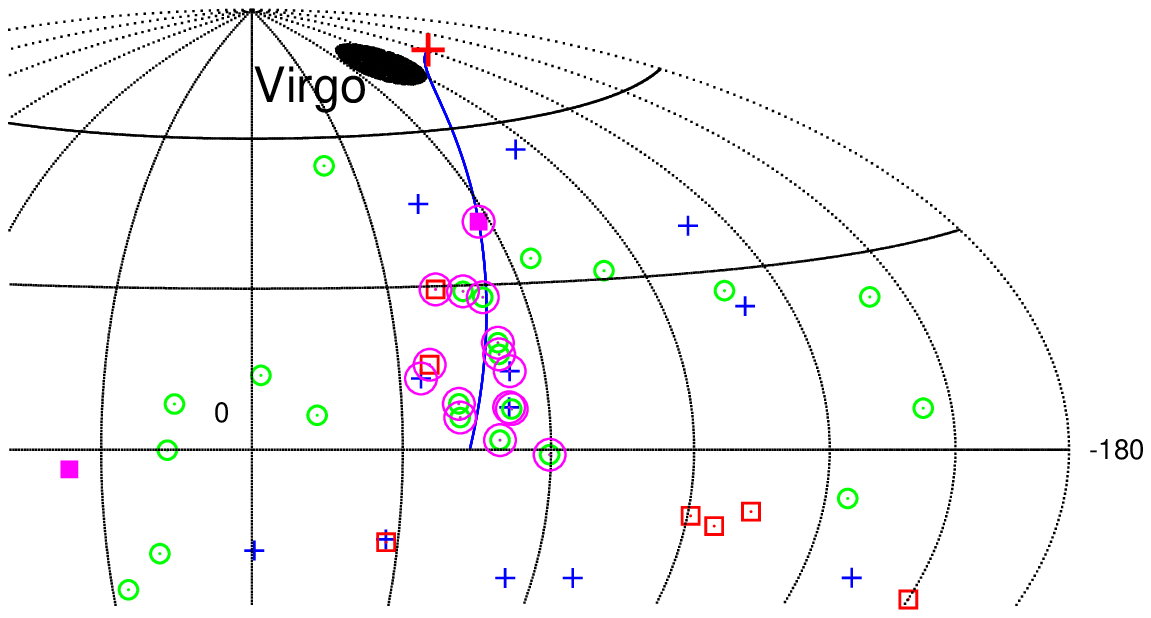}
\includegraphics[width=0.49\textwidth]{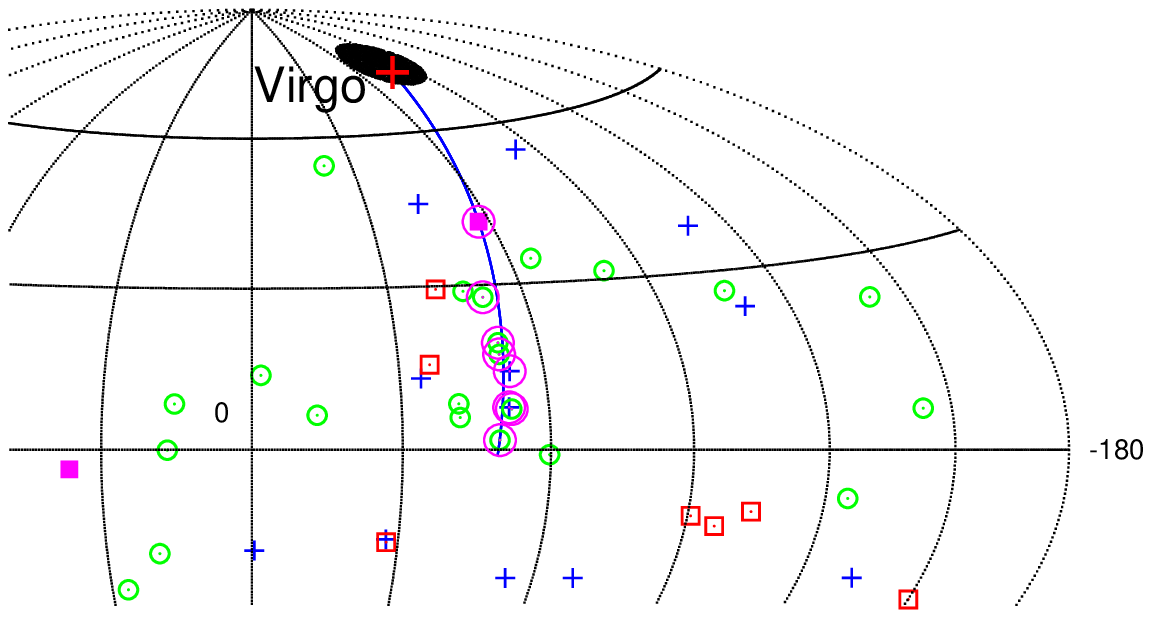}
\end{center}
\caption{Sector with: \textbf{Upper panel:} $\Theta=60^{\circ}$. There are 15 events in the sector; \textbf{Lower panel:} $\Theta=20^{\circ}$. There are 8 events in the sector. Same key as in Fig.~\ref{Sector40degrees}. The source is reconstructed at ($l \simeq -122^{\circ}$, $b \simeq 74.8^{\circ}$) for $\Theta=60^{\circ}$, and at ($l \simeq -78.4^{\circ}$, $b \simeq 72.4^{\circ}$) for $\Theta=20^{\circ}$.}
\label{Sector60and20degrees}
\end{figure}

\begin{itemize}
 \item \textit{$\Theta=60^{\circ}$ -see Fig.~\ref{Sector60and20degrees} (upper panel):} when one considers this larger sector angle, one selects two more events in the overdense region, compared to the case $\Theta=40^{\circ}$. As visible in Fig.~\ref{Sector60and20degrees}, all the cosmic rays which belong to the overdense region are taken into account in this configuration. $\mathcal{N}=15$ and $C_{\min}=0.6$ (because $Corr(X',1/E) \simeq 0.61$). The reconstructed source is located at $\simeq11.8^{\circ}$ from M87, at ($l \simeq -122^{\circ}$, $b \simeq 74.8^{\circ}$). The results are slightly worse than for $\Theta=40^{\circ}$ which is the ``true'' minimum for the considered sets of parameters, but they are not very far. For this case, the probabilities introduced above become: P$_{feature}\simeq1.1\times10^{-2}$ (respectively 6150 and 503901 cases out of $4.7\times10^{7}$ before and after the penalization), P$_{feature,d(M87)<10^{\circ}}\simeq6\times10^{-5}$ (resp. 30 and 2658), P$_{d(M87)<10^{\circ}}\simeq5\times10^{-3}$ (0.5\%).
 \item \textit{$\Theta=20^{\circ}$ -see Fig.~\ref{Sector60and20degrees} (lower panel):} we discuss this smaller sector, because the source is reconstructed much closer to M87. However, it only takes into account 8 points from the overdense region. The reconstructed source is located at $\simeq2.2^{\circ}$ from M87, at ($l \simeq -78.4^{\circ}$, $b \simeq 72.4^{\circ}$), which is located in the Virgo cluster. $\mathcal{N}=8$ and $C_{\min}=0.7$ (because $Corr(X',1/E) \simeq 0.78$). These 8 events belong to a ``filamentary'' structure which is a denser sub-region of the ``right'' part of the overdense region. Knowing if this filamentary structure is the real image of the Virgo cluster, instead of the whole overdense region, is beyond the scope of what one could currently say. We compute the same probabilities as above, except that we take into account sources reconstructed at distances below 3$^{\circ}$ from M87, instead of 10$^{\circ}$. Here, P$_{feature}\simeq2.1\times10^{-1}$ (respectively 128469 and 9987251 cases out of $4.7\times10^{7}$ before and after the penalization), P$_{feature,d(M87)<3^{\circ}}\simeq9\times10^{-5}$ (resp. 53 and 4368), P$_{d(M87)<3^{\circ}}=\mbox{P}_{feature,d(M87)<3^{\circ}}/\mbox{P}_{feature}\simeq4\times10^{-4}$. The probability to have in the background such a filamentary structure of $\geq8$ events and $Corr(X',1/E) \geq 0.7$, with $\mathcal{D}=39^{\circ} \times 10^{20}$\,eV, is much higher than the probability to have the features with $\geq13\,\mbox{or}\,15$ events for $\Theta=40^{\circ}\,\mbox{and}\,60^{\circ}$. However, as shows the value of P$_{d(M87)<3^{\circ}}$, once one has such a feature, the probability to reconstruct the source at less than 3$^{\circ}$ to M87 is naturally around 10 times lower than the probability to reconstruct it at less than 10$^{\circ}$ -see the case of $\Theta=40^{\circ}$.
\end{itemize}

\begin{figure}
\begin{center}
\includegraphics[width=0.49\textwidth]{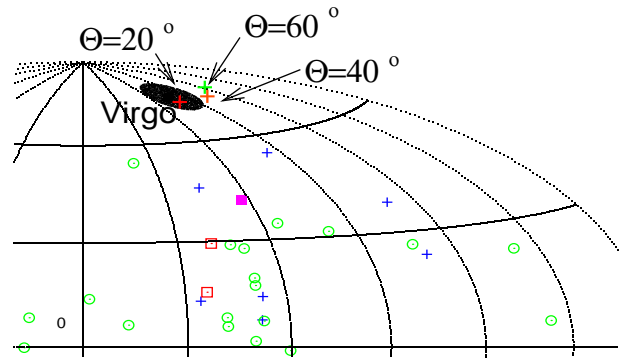}
\end{center}
\caption{Positions of the reconstructed sources on the celestial sphere for the three considered sector opening angles: $\Theta=20\mbox{, }40\mbox{, and }60^{\circ}$. Same key as in previous Figures.}
\label{PosRecSource}
\end{figure}

Fig.~\ref{PosRecSource} shows on the same sky map the three positions of the reconstructed sources for the cases: $\Theta=20^{\circ}$, $40^{\circ}$ and $60^{\circ}$. The larger the sector angle, the further from M87 the reconstructed source. This could simply mean that the events are not deflected along a straight line on the sphere. This would not be surprising for deflections of several tens of degrees, even for ``enlarged proton-like'' images~\cite{Giacinti:2010dk}. On the contrary, this might mean that only the events in the sector $\Theta=20^{\circ}$ come from Virgo. However, having such a thin linear filamentary structure on the sky for such large deflections would be hard to realize. Let us note that only one event has come in this sector in the second Auger data set of $69-27=42$ events.

No noteworthy feature is found in the data when applying the method presented in Section~\ref{Method} to the two other events with energies above 10$^{20}$\,eV (115 and 123\,EeV events).


\section{Cross-check with a blind-like analysis}
\label{BlindAnalysis}

We shall now check the result of the previous section by doing a ``blind-like'' analysis. It consists in choosing the best sector for the first data set released by the Pierre Auger Collaboration, which contains 27 events. We take the sector for which the measured signal (number of events and correlation coefficient $Corr(1/E,X')$) has the lowest probability to be reproduced by the background. The method used to define this sector can contain as many parameters as needed. One fixes the best parameters and then does not have to penalize over them.

The second step consists in analysing the ``newer'' $69-27=42$ events of the second Auger data set with this fixed ``best sector''.

\begin{figure}
\centering
\includegraphics[width=0.49\textwidth]{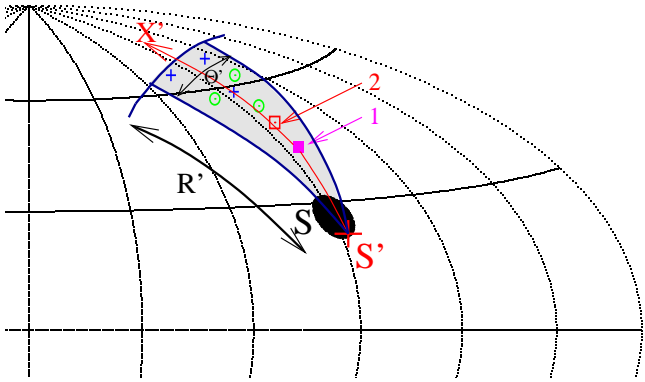}
\caption{Generic way to define the second sector which is used in the study of Section~\ref{BlindAnalysis}. Same key as in Fig.~\ref{Fig_Method}. $R'$ and $\Theta'$ defined in the text.}
\label{Fig_Method2}
\end{figure}

For proton sources or ``enlarged proton-like'' images of nuclei sources, one can expect that events are roughly deflected in a sector which vertex is theoretically located at the source position (see Ref.~\cite{Giacinti:2009fy} for a full explanation). Nevertheless, the source position is \textit{a priori} unknown. Since the highest energy event is near the source position for proton sources, one can take the highest energy event as the origin of the sector in this case. For UHE heavy nuclei deflected in the GMF, the distance between the source and its highest energy events is usually estimated to be of the order of a few tens of degrees. Taking, as in the previous section (see Fig.~\ref{Fig_Method}), the highest energy event as the origin of the sector gives good results in practice. However, one may miss in some cases one or two source events at the very border of the sector, notably in the regions near the highest energy event.

In this section, we can take this point into account for the method used to define the ``best sector''. This improved method contains a fifth parameter, named $\Theta'$ below. It starts in the same way as the method of Section~\ref{Data}: the vertex of a first sector is set on the highest energy event and the source position is reconstructed as previously. The only difference is that the reconstructed source position is now regarded as the vertex of a second sector as shown in Fig.~\ref{Fig_Method2}. The highest energy event, denoted by ``1'' in Fig.~\ref{Fig_Method2}, defines the central axis of this second sector. It has an opening angle $\Theta'$ and extends up to $R'=\mathcal{D}/$(55\,EeV). It is highlighted in grey in Fig.~\ref{Fig_Method2}. In this section, it is the sector in which one counts both $\mathcal{N}$ and $Corr(1/E,X')$. Its geometry is more adapted to grab the events of an ``enlarged proton-like'' image of a source located near its vertex.

\begin{figure}
\centering
\includegraphics[width=0.48\textwidth]{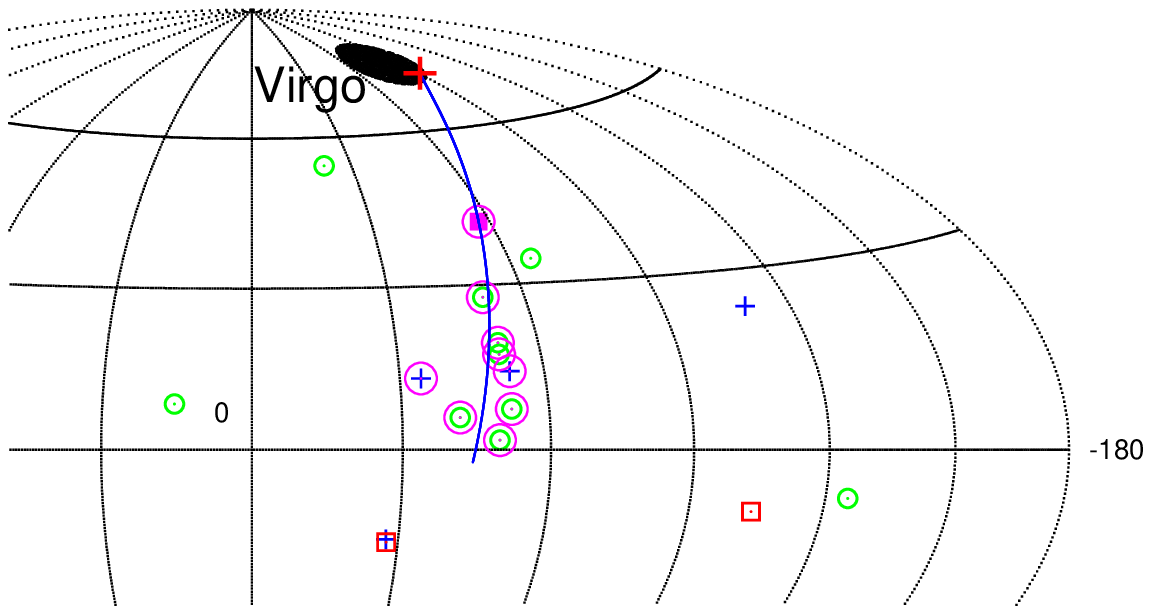}
\includegraphics[width=0.48\textwidth]{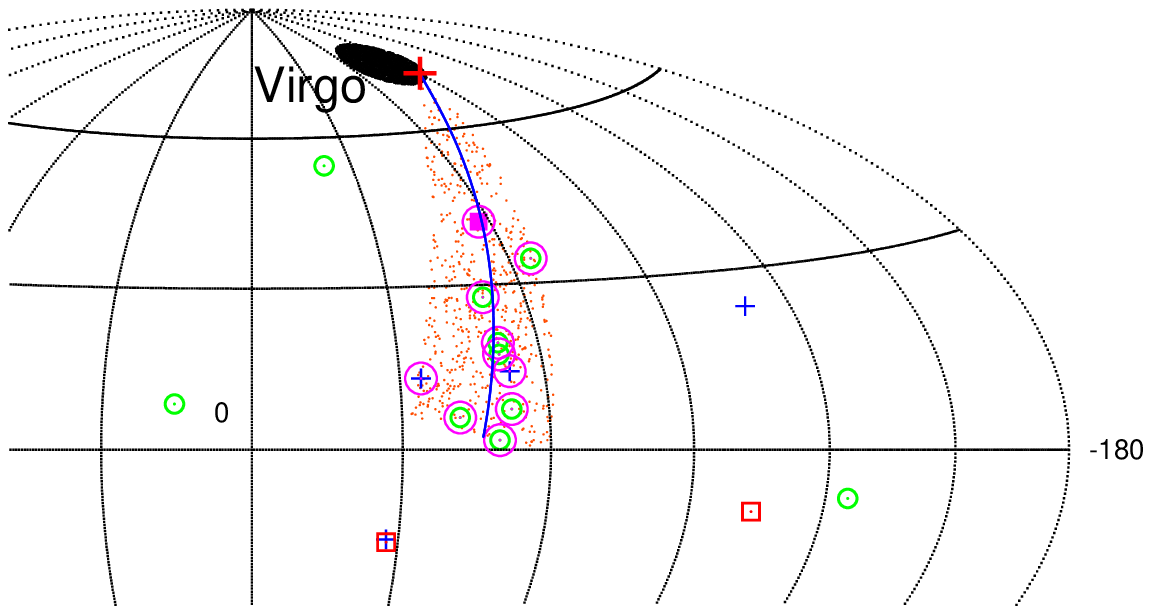}
\includegraphics[width=0.48\textwidth]{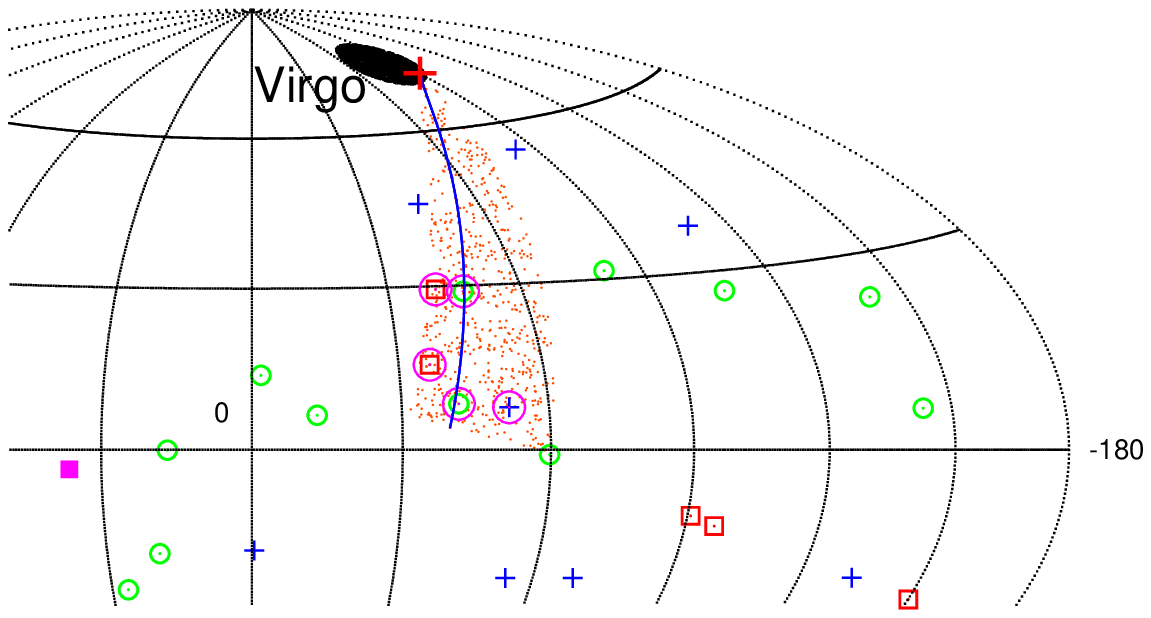}
\caption{Blind-like analysis conducted with the two consecutive data sets of Auger (27 and $69-27=42$ events), plotted here in Galactic coordinates. \textbf{Upper panel:} First sector used for the reconstruction of the source position with the data set of 27 events. Red cross for the reconstructed source position; \textbf{Middle panel:} The ``best sector'' for the first 27 events. It starts from the red cross and is highlighted in orange. It contains 10 out of the 27 events. The correlation coefficient between $1/E$ and $X'$ is $Corr(1/E,X') \simeq 0.73$; \textbf{Lower panel:} The ``best sector'' for the first 27 events (highlighted in orange) applied to the newer $69-27=42$ events. It contains 5 events, and the correlation coefficient is $\simeq 0.38$. Same key as in Fig.~\ref{Sector40degrees}. On each panel, events located in the sectors are surrounded by magenta circles. See text for details on the ``best sector''.}
\label{BLA}
\end{figure}

For the 27 events data set, the best first sector is obtained for $\mathcal{D}\sim41^{\circ} \times 10^{20}$\,eV, $\Theta \simeq 40^{\circ}$, and contains 9 events (including the highest energy event). The correlation coefficient is $Corr(1/E,X') \simeq 0.77$. The source is reconstructed at ($l \simeq -90.6^{\circ}$, $b \simeq 71.3^{\circ}$), at $\simeq 5.3^{\circ}$ from M87. See upper panel of Fig.~\ref{BLA}. The events in the sector are surrounded by magenta circles and the reconstructed source position is denoted by the red cross. The probability to have with some background at least 9 events with $Corr(1/E,X') \geq 0.77$, for $\mathcal{D}=41^{\circ} \times 10^{20}$\,eV and $\Theta=40^{\circ}$, is P$_{1}\sim 6\times10^{-6}$.

The vertex of the second sector coincides with the position of the reconstructed source. The best values for this sector are $\mathcal{D}\simeq41^{\circ} \times 10^{20}$\,eV, $\Theta' \simeq 30^{\circ}$. For the 27 events data set, it contains $\mathcal{N}=10$ events (including the 142\,EeV highest energy event) and the correlation coefficient is $Corr(1/E,X') \simeq 0.73$. We now fix this sector which represents the ``best sector'' for the first Auger data set of 27 events. See middle panel of Fig.~\ref{BLA}, where it is highlighted in orange. The probability to have at least 10 events and $Corr(1/E,X') \geq 0.73$ in this fixed sector with some background made of 27 events, and following the Auger exposure and spectrum, is P$_{2}\sim10^{-8}$.

We can now apply this ``best sector'' to the second Auger data set of $69-27=42$ events. We find inside $\mathcal{N}=5$ events, with a correlation coefficient $Corr(1/E,X') \simeq 0.38$. See lower panel of Fig.~\ref{BLA}. For the first data set of 27 events, the significance was higher than for the data set of 42 events. We finally confront this result with some random background made of 42 events following the Auger exposure and spectrum. We find that the probability to have with the background $\mathcal{N}\geq5$ events and $Corr(1/E,X') \geq 0.38$ in the ``best sector'' is equal to \textbf{P$_{BLA} \simeq 1.8 \times 10^{-2} \sim$\,2\%}.

Let us note that if one adds or removes by hand one border point in this ``best sector'', the probability P$_{BLA}$ can non-negligibly vary and take for instance values as $\sim1\%$ or $\sim6\%$. This is due to the small number of points. The value of P$_{BLA}$ has to be checked in the future with more statistics. The order of magnitude of 2\% is compatible with the result found in the previous section (0.4\%).

The analysis conducted in this section is not a real blind analysis, since it is done \textit{a posteriori}. A real blind analysis can start now, by fixing the best sector for the 69 Auger events and looking at the future data. We should expect the number of events to increase in this sector, preferably more rapidly in average than elsewhere. However, the correlation coefficient $Corr(X',1/E)$ will not necessarily increase.


\section{Discussion}
\label{Discussion}


The results of the two previous sections show that, in the Auger data, the 142\,EeV event and the events in the Cen A region overdensity are compatible with an emission from the Virgo cluster.

As discussed in Ref.~\cite{DSPaper}, the Cen A region events may be emitted by Virgo even if they are protons. If extragalactic magnetic fields (EGMFs) are as large as in Refs.~\cite{SiglEGMF,Sigl:2004gi}, UHE protons could experience deflections as large as several tens of degrees.

On the contrary, if EGMFs are as low as in Refs.~\cite{Dolag:2003ra,DolagEGMF}, this would favour a heavy nuclei origin. Outside clusters, nuclei would not experience substantial deflections in the EGMF. They would be mostly deflected by the GMF, with such typical values.

In the Galactic disk, the GMF is mostly parallel to the plane of the disk~\cite{Heiles,Han:2006ci}. In the existing GMF models, the field in the halo is also assumed to be parallel to the disk~\cite{Jansson:2009ip}. In this case, nuclei from high latitude sources are approximately shifted along lines of equal Galactic longitudes. This would be consistent with the possibility that the considered events come from Virgo. The 142\,EeV event, Cen A region and Virgo nearly have same longitudes. For the GMF structure, the exception is near the Galactic center, where a dipolar contribution may create a substantial component of field perpendicular to the Galactic plane.

\begin{figure}
\begin{center}
\includegraphics[width=0.49\textwidth]{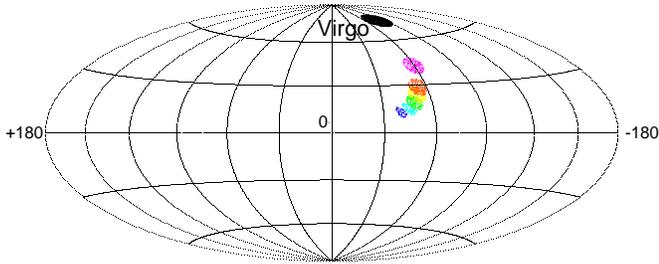}
\end{center}
\caption{Image in Galactic coordinates of UHE iron nuclei emitted by the Virgo cluster and deflected in the regular GMF model which parameters are discussed in the text. Adding to these computations the deflections due to the turbulent GMF would spread the lower energy events in the whole Cen A region. The black disk represents the Virgo galaxy cluster. Shaded areas represent the arrival directions at Earth of cosmic rays with given energies. Dark blue stands for 60\,EeV nuclei, light blue for 70\,EeV, green for 80\,EeV, yellow for 90\,EeV, orange for 100\,EeV, and magenta for 140\,EeV.}
\label{VirgoImage}
\end{figure}

As shown in Ref.~\cite{virgoMF}, even if there is only one source in the Virgo galaxy cluster, the cluster should shine as a whole due to substantial deflections of UHECRs in the magnetic fields inside. This means that the Virgo cluster can be considered as an extended source, basically a $\simeq5^{\circ}$ radius disk on the celestial sphere. Therefore, in Fig.~\ref{VirgoImage} we model Virgo as a $5^{\circ}$ radius source.

We used the numerical code of Ref.~\cite{Giacinti:2010dk} to deflect iron nuclei, with 60\,EeV to 140\,EeV energies, in different GMF models. We found that we can model the Auger data with several modified and reshaped versions of existing theoretical GMF models. For example, Figure~\ref{VirgoImage} shows such a modeling of the Auger data. In this figure, only deflections in the regular GMF are considered. If one adds the deflections in the turbulent field, the lower energy events would be spread in the whole Cen A region. The image would then look like the considered feature in the Auger data. The black disk represents the Virgo galaxy cluster, while shaded areas show the arrival directions of UHE iron emitted by Virgo, with a given energy. The dark blue region corresponds to 60\,EeV iron nuclei, light blue to 70\,EeV, green to 80\,EeV, yellow to 90\,EeV, orange to 100\,EeV, and magenta to 140\,EeV.

With notations and names introduced in Ref.~\cite{Giacinti:2010dk}, the regular GMF model used for Figure~\ref{VirgoImage} is comparable to a ``Sun08 model'' with modified parameters, and with a dipolar contribution similar to the ``Prouza and Smida'' model, with $\mu_{D}=30\,\mu$G$\cdot$kpc$^{3}$. For the halo, we took $B_{T0}=0.8\,\mu$G, $h_{T}=2\,$kpc, $w_{T,in}=1.5\,$kpc, $w_{T,out}=2\,$kpc and $r_{T0}=8.5\,$kpc. For the disk, $B_{0}=B_{c}=2\,\mu$G, $r_{c}=5\,$kpc, $r_{0}=10\,$kpc, $z_{0}=0.2\,$kpc and $p=35^{\circ}$. We also added a $-60^{\circ}$ pitch angle in the halo. This specific configuration is not the only one which could lead to an image compatible with the data. Hence, it should not be considered as a prediction on the configuration of the GMF, or on its extension or strengths in the disk and the halo. It however proves that some configurations of the GMF are compatible with the interpretation of the Auger data discussed in this paper.

The relatively low value for the correlation coefficient $Corr(1/E,X')$ computed in Section~\ref{Data} can be notably due to the spread of arrival directions in the Cen A region due to the turbulent GMF. Such a small energy ordering in the image is consistent with what one can expect in case of heavy nuclei sources.

If Virgo will be confirmed in the future to be a UHE nuclei source, it will put strong constraints on the Galactic and extragalactic magnetic fields. First, deflections in the EGMF would be small compared to the deflections in the GMF (except in the case of proton primaries, discussed above). Second, the shift of the 142\,EeV event would allow to give an immediate estimate of typical deflections of cosmic rays in the Northern halo of the GMF. Third, the small scattering of arrival directions of cosmic rays around the thin linear structure they would have had in the regular field alone (see for example Fig.~\ref{VirgoImage}), would imply that the deflections in the GMF are mostly dominated by the regular field contribution. The deflections in the turbulent field would be small enough not to destroy the image at the lowest energies.

The confirmation that the events in the Cen A region have been emitted by Virgo would put additional constraints on the regular GMF:

First, we noticed in Ref.~\cite{Giacinti:2010dk} that large dipolar or toroidal contributions to the GMF can make UHE nuclei sources at the Galactic poles invisible. The data from future radio experiments will enable us to have a better knowledge on the strength and extensions of these components. If the Virgo origin of the Cen A region events is proved, it would independently bring strong constraints on the maximum contributions of the dipolar and toroidal components.

Second, it would also put tight constraints on the disk field. The lower energy part of the image is in the Cen A region, which is not far from two stronger field regions: both the Galactic center direction and the Galactic plane. The computations of the images of UHE iron from Virgo, deflected in different GMF models, show that the shape of the image is very sensitive to the exact GMF configuration. For heavy nuclei, the influence of the disk field starts to be substantial in the region $b\la30^{\circ}-40^{\circ}$, if its typical height extension is non-negligible (for example for $z_{0}\ga1\,$kpc). The alignment of the image along constant $l$ ($l\sim-60^{\circ}$ to $-30^{\circ}$) from Virgo to the plane would enable one to exclude several configurations of the field.

Third, in our modeling, pitch angles from $\sim-40^{\circ}$ to $\sim-60^{\circ}$ in the halo reproduced well the Auger data. This may suggest a non-negligible pitch angle in the halo field.

This would also lead to a better understanding and tighter constraints on UHECR sources. Only a few extreme astrophysical objects can accelerate particles to such energies~\cite{Hillas:1985is,Ptitsyna:2008zs}. Let us note that the Auger UHECR flux in the Cen A region is of the same order as the gamma ray flux from M87, measured by HEGRA~\cite{Aharonian:2003tr} or by H.E.S.S., MAGIC and VERITAS~\cite{Wagner:2009rw,Acciari:2010uh}. Being able to discriminate between the scenarii of an image created by one source or by several different sources in the Virgo cluster would require much more statistics than currently available.

Reference~\cite{Lemoine:2009pw} adds an important constraint on the Cen~A region overdensity. Due to their equal rigidities, 60 - 80\,EeV iron nuclei would be deflected as 2 - 3\,EeV protons, whatever the strengths and structures of the EGMF and GMF are. Hence, if UHECR sources accelerate both nuclei and protons and if this Cen A overdensity is made of heavy nuclei, one should expect protons at a $Z$ times lower energy, exactly in the same region. The Auger flux is however compatible with isotropy at low energies, 2 - 3\,EeV~\cite{Lemoine:2009pw}. So if these events are nuclei, the results of Ref.~\cite{Lemoine:2009pw} imply either that the source spectrum is harder than a $1/E^2$ spectrum or that the ratio of accelerated protons to nuclei in the source is not more than one to one. The source(s) in Virgo can have a very hard spectrum as, for example, in the model presented in Ref.~\cite{Neronov:2009zz}. A fraction of the emitted nuclei are destroyed on their way to the observer and produce lighter nuclei. The events in the Cen A region would correspond to nuclei with a Lorentz factor of $\Gamma\sim10^{9}$. As shown in figure 1 (left panel) of Ref.~\cite{Allard:2005ha}, such nuclei have a mean free path of the order of 100\,Mpc, which is much larger than the distance to Virgo. Besides, the propagation of nuclei in galaxy clusters has been studied in Ref.~\cite{Kotera:2009ms}. The authors find that the results mostly depend on the source position, as well as on the strength and profile of the magnetic field in the cluster. Therefore, if most of the emitted nuclei manage to escape the cluster, only a small fraction will be destroyed and enhance the light nuclei flux at lower energies. Let us note that for energies above a few times $10^{20}$\,eV, iron and intermediate nuclei have a mean free path smaller than the distance to Virgo (see figure 1 of Ref.~\cite{Allard:2005ha}). Therefore, the maximum acceleration energy of the source should not be too high. Otherwise, an additional flux of protons due to the disintegration of its highest energy iron nuclei could be seen at lower energies in the data. Thus the confirmation of a nuclei source in the Virgo cluster would put interesting constraints on acceleration mechanisms, on the composition of particles accelerated in the source and on physical conditions in the Virgo cluster.


We show below that it will be possible to confirm or rule it out in the future, when more experimental data will be available. At the same time, we present other possibilities which can explain the present Auger data.

One can expect that Auger South experiment will triple its statistics during its lifetime. It will confirm if the overdensity in the Cen A region is not a statistical fluctuation. Auger North experiment would also be useful to check if there are comparable features in the Northern hemisphere.

If the Cen A overdensity really exists, there are two cases. It is either due to protons, or to nuclei.

If the results of HiRes on the composition of primaries are correct, the events in the Cen~A region are protons. If deflections in the EGMF are as low as in Refs.~\cite{Dolag:2003ra,DolagEGMF}, most of the Cen~A region events should be protons emitted by the Centaurus galaxy cluster~\cite{AugerPaper}. Refs~\cite{Gorbunov:2008ef,Gorbunov:2007ja} argue that this explanation would be challenged by the lack of events in the Virgo direction. Even if the Auger exposure is smaller in the direction to Virgo than in the direction to Centaurus, Virgo is closer to our Galaxy and one should statistically see at least a few protons coming from its direction. The Centaurus cluster lies behind Cen~A. Cen~A has been regarded as a potential source of UHE protons for a long time~\cite{Cavallo,Romero:1995tn,Anchordoqui:2001nt}. For such a composition, it may be the source of two cosmic rays in this direction~\cite{Abraham:2007si,Abraham:2010mj,Kachelriess:2008qx}.

If the results of Auger composition studies are correct, the events in the Cen~A region are nuclei. In this case, there are currently three main explanations. Either the UHECR deflections in the GMF and EGMF are large enough to prevent the identification of nuclei sources, or at least one source can be detected. In the first case, the higher flux of UHECR in this direction may just be due to magnetic lensing effects. Such effects have been studied in a particular example of lens geometry by Ref.~\cite{Battaner:2010bd}. They have been quantified for UHE iron propagated in models of the regular and turbulent GMF by Refs.~\cite{Giacinti:2010dk,TurbGMFiron}. In the second case, there are two possibilities. First, these nuclei may have been emitted by Virgo and shifted in the GMF, as studied in the present paper. Second, they may have been emitted by Cen A~\cite{Gorbunov:2008ef}. However, there are arguments that Cen A may not be powerful enough to accelerate cosmic rays to such extreme energies~\cite{Casse:2001be,Kotera:2008ae,Trondheim}.

There are two requirements to confirm that Virgo is the source. One must prove that both the 142\,EeV event and the Cen A region events are connected to Virgo.

Proving the link between Virgo and the highest energy event, should be easier than for the Cen A region events. If other events with $E>100$\,EeV come in the region of the 142\,EeV event, and are located at places approximately compatible with a collective emission by the Virgo cluster, one could prove their Virgo origin. If Virgo is the source, Auger may detect such an event during its lifetime. It would be a hint. However, the final confirmation should only be given by the next generation of UHECR experiments.

Checking the link between the Cen A region and Virgo will require data from the next generation of UHECR experiments, such as JEM-EUSO. A confirmation of the Virgo origin of the highest energy event will not automatically imply that the Cen A region events have been emitted by Virgo. The 142\,EeV event may be a nucleus from Virgo and be disconnected from the Cen A region events which may have another source. Finding some events in between, with more intermediate energies would be particularly valuable to validate that the Cen A region events come from Virgo. JEM-EUSO will have one order magnitude more data. It is expected to reach in few years of observation an exposure of 10$^{6}$km$^{2}\cdot$sr$\cdot$yr~\cite{MedinaTanco:2009tp}. It will detect more than 1000 events at such energies. If the events are nuclei from Virgo, the link between the events with the highest energies and the Cen A lower energy overdensity should become clearer, suggesting a common origin.

On the contrary, if the Cen A region events are protons from the Centaurus cluster, JEM-EUSO must be able to see the first characteristic individual images of UHE proton sources.

\section{Conclusions and Perspectives}
\label{Conclusion}

In this paper, we have proposed a new method to search for images of UHE heavy nuclei sources in the data, on top of background.

We have pointed out that for some GMF configurations, and for some source positions on the sky, one can still have roughly ``enlarged proton-like'' images at energies $E\ga60\,$EeV, even for iron primaries. In this case, one can detect the sources if they are bright enough and reconstruct their positions on the celestial sphere with the method we presented in Section~\ref{Method}.

Detecting a source in this way would however not always be possible, because images of iron sources can often exhibit more complicated patterns. In general, a much better knowledge on the GMF (and EGMF) than currently available is necessary to detect UHE heavy nuclei sources and reconstruct their positions on the sky. Future radio telescopes, such as SKA and its precursor LOFAR, will increase the number of rotation measures on the sky by a few orders of magnitude. This will, for example, enable us to know the geometry of the regular GMF in the halo and in the disk, as well as the turbulent GMF strength and properties~\cite{Gaensler:2004gk,Beck:2004gq}.

In Section~\ref{Data}, we checked if one could find such a case in the data recently published by the Pierre Auger Collaboration~\cite{AugerPaper}. We found that the Cen A region overdensity and the 142\,EeV event may be the image in UHE nuclei of the Virgo cluster, deflected by the GMF. With our method, the associated source position is reconstructed near the Virgo cluster, at only $\simeq8.5^{\circ}$ from M87. This indicates that these events are compatible with a common origin from the Virgo cluster. The probability to have such a feature in some random background and reconstruct the source at less than $10^{\circ}$ from M87 is about $3\times10^{-5}$. If one assumes that the Cen A region overdensity is due to another reason, and that the 142\,EeV event appeared by chance near this region, the probability to reconstruct with our method the source at less than 10$^{\circ}$ from M87 is $\simeq0.4\%$. In Section~\ref{BlindAnalysis}, we performed a ``blind-like'' analysis, by dividing the Auger data set in two parts: the first 27 Auger events and the $69-27=42$ remaining events. We determined for the 27 first events the ``sector'' on the celestial sphere for which the probability to reproduce the data with some random background was the lowest. We fixed it and analyzed the $69-27=42$ newer events with this sector. The probability to have by chance the signal detected in the sector for the second data set is $\sim2\%$.

If future data confirm that the feature discussed in this paper is due to UHE nuclei from Virgo, it would lead to significant improvements in our knowledge both on the cosmic magnetic fields and on the UHECR acceleration mechanisms. It would imply that deflections in the extragalactic magnetic fields are negligible compared to deflections in the Galactic magnetic field. Moreover, deflections would be dominated by the regular GMF, which structure and strength would be better constrained.

Thus, we have presented here both a new method to look for ultra-high energy nuclei sources on top of background, and a new and consistent way to interpret the Auger data. We have found that one (or several) ultra-high energy nuclei source(s) in the Virgo galaxy cluster could explain both the composition and the anisotropy in the Auger data. However, a better knowledge of the Galactic magnetic field than currently available, or more UHECR data are still needed to confirm or rule out this possibility.

\section*{Acknowledgments}

We would like to thank Michael Kachelrie$\ss$ for useful comments on this work.

\end{document}